\documentclass{llncs}

\usepackage{bytefield}
\usepackage{listings}
\usepackage{booktabs}
\usepackage{array}
\usepackage{pgfplotstable}    
\usepackage{textcomp}
\usepackage{mathtools}
\usepackage{paralist}
\usepackage{algorithm}
\usepackage{algpseudocode}
\usepackage{environ}
\usetikzlibrary{patterns}
\pgfplotsset{compat=1.11}

\lstdefinestyle{mystyle}
{
  language=Fortran,
    frame=single,
    tabsize=4,
    showspaces=false,
    showstringspaces=false,
    keywordstyle=\color{blue}\bfseries,
    keywordstyle = [2]{\color{blue}},
    keywordstyle = [3]{\color{black}\bfseries},
    morekeywords = [2]{string},
    morekeywords = [3]{EntityInfo},
    breaklines=true,
    escapeinside={(*}{*)}
}

\NewEnviron{smallequation}{%
    \begin{equation}
    \scalebox{0.75}{$\BODY$}
    \end{equation}
    }

\begin{document}\sloppy
\mainmatter 

\title{Internet of Entities (IoE): a Blockchain-based Distributed Paradigm to Security}

\author{Roberto Saia}
\authorrunning{Roberto Saia}
\institute{Department of Mathematics and Computer Science\\
University of Cagliari, Via Ospedale 72 - 09124 Cagliari, Italy\\
\email{roberto.saia@unica.it}
}
\maketitle

\begin{abstract}
The exponential growth of wireless-based solutions, such as those related to the mobile smart devices (e.g., smart-phones and tablets) and Internet of Things (IoT) devices, has lead to countless advantages in every area of our society.
Such a scenario has transformed the world a few decades back, dominated by latency, into a new world based on an efficient real-time interaction paradigm.
Recently, cryptocurrency have contributed to this technological revolution, the fulcrum of which are a decentralization model and a certification function offered by the so-called blockchain infrastructure, which make it possible to certify the financial transactions, anonymously.
However, it should be observed how this challenging scenario has generated new security problems directly related to the involved new technologies (e.g., e-commerce frauds, mobile bot-net attacks, blockchain DoS attacks, cryptocurrency scams, etc.).
In this context, we can acknowledge that the scientific community efforts are usually oriented toward specific solutions, instead to exploit all the available technologies, synergistically, in order to define more efficient security paradigms.
This paper aims to indicate a possible approach able to improve the security of people and things by introducing a novel blockchain-based distributed paradigm to security defined Internet of Entities (IoE).
It represents an effective mechanism for the localization of people and things, which exploits both the huge number of existing wireless-based devices and the blockchain-based distributed ledger technology, overcoming the limits of traditional localization approaches, but without jeopardizing the user privacy. 
Its operation is based on two core elements with interchangeable roles, entities and trackers, which can be very common elements such as smart-phones, tablets, and IoT devices, and its implementation requires minimal efforts thanks to the existing infrastructures and devices.
The possibility of including further information to those of localization, such as those generated by device sensors, gives rise to a novel and widely exploitable data environment, whose applications can be extended to contexts different from that of the localization of people and things, e.g., eHealth, Smart Cities, and so on.
\keywords{Internet $\cdot$ Internet of Things $\cdot$ Internet of Entities $\cdot$  Mobile Network $\cdot$ Blockchain $\cdot$ Distributed Ledger $\cdot$ Localization $\cdot$ Security}
\end{abstract}

\section{Introduction}
\label{IntroductionSection}
The meaning of the personal security is day after day closer to that of the data security. 
This is given by the growing number of activities related with everyday life, which are somehow performed in a virtual way (e.g., requests for documents, job applications, purchases, and so on).

Such a scenario has been further revolutionized by the decentralized paradigm introduced with the advent of the \textit{Bitcoin}~\cite{Bonneau15ieeesp} cryptocurrency, which has traced a new way to exchange currency.
A synergistic combination of \textit{security} and \textit{anonymity} stands at the base of its success, since this paradigm allows the users to exchange currency without the need to involve trusted authorities as intermediates.

The strategy behind this revolutionary way to operate is mainly based on a digital signature scheme, which is combined with the effort needed to solve a quite hard mathematical problem, but the real fulcrum of this mechanism is an immutable public ledger where all the transactions are recorded.
It is implemented on the so-called \textit{blockchain-based} infrastructure by exploiting a distributed consensus protocol that operate in a peer-to-peer network~\cite{nakamoto2008bitcoin}.

The idea on which the proposed \textit{IoE} paradigm revolves is the exploitation of the \textit{wireless-based} ecosystem, where some existing devices (hereinafter referred to as \textit{trackers}) are used in order to track the activity of other devices associated to people or things (hereinafter referred to as \textit{entities}), registering a series of immutable information about the latter ones by using the features offered by a \textit{blockchain-based distributed ledger}. 
This idea relies on what affirmed by several authoritative studies, which indicate that by the end of this decade the number of \textit{smart-phones} and \textit{tablets} will be about \textit{7.3} billion of units~\cite{pop2017harvesting}, as well as the number of \textit{IoT} devices, which will be between \textit{20} and \textit{50} billion by \textit{2020}~\cite{reyna2018blockchain}.

\begin{figure}[!]
\centering
\scriptsize
\includegraphics[width=0.8\textwidth]{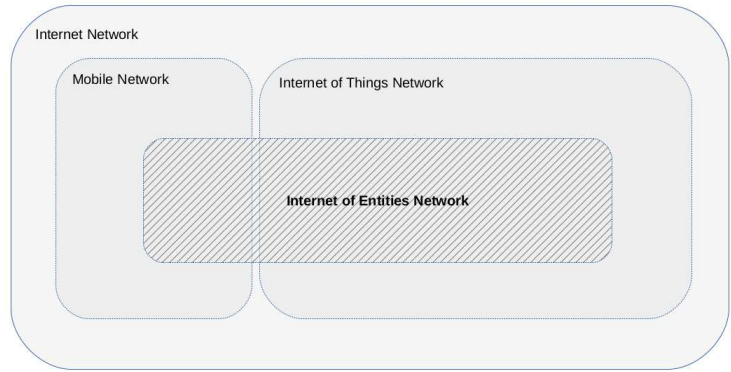}
\caption{$IoE\ Placement$}
\label{IoePlacementGraph}
\end{figure}

Although in a rather coarse manner, Figure~\ref{IoePlacementGraph} shows the placement of the proposed \textit{IoE} paradigm, with respect to the other already existing \textit{wireless-based} paradigms: it is straddled on their operative areas.  

The implementation of such a paradigm can be made by adding simply functionalities to the existing devices used as \textit{trackers} (\textit{IoT}, \textit{Smart-phones}, etc), since we only need to append few \textit{entity} data (i.e., \textit{unique identifier} and \textit{sensors data}) with few \textit{tracker} data (e.g., \textit{time-stamp}, \textit{geographic location}, \textit{sensors data}, etc.) and sent them to a \textit{blockchain-based} distribute ledger. 
It should be observed that in case of devices such as \textit{smart-phones} and \textit{tablets}, such a operation can be performed in a quite transparent way, by installing a simple application, while for the \textit{IoT} devices, it can be done by performing a software update.

About the \textit{entity}-side of this scenario, an interesting aspect related to the \textit{IoE} paradigm is its capability to exploit as \textit{entities} both custom devices (e.g., light wearable devices) or existing widespread devices (e.g., \textit{smart-phones}).
In addition, the \textit{IoE} paradigm operates anonymously, since only the \textit{entity} owner can associate its unique identifier to the registration performed on the remote ledger through the \textit{trackers}.
The inclusion, when it is applicable, of one or more \textit{neighbor entities} (i.e., those detected by the \textit{tracker} near the \textit{entity} within a given \textit{time-frame}) offers an additional tracing opportunity, since it allows us to reconstruct an \textit{entity} activity in a wide manner, without jeopardize the anonymity of the involved \textit{neighbor entities}.

In should be observed how in addition to the domain strictly related to security, such as that proposed in this paper, there are other areas where the \textit{IoE} paradigm can be profitably exploited (e.g., \textit{eHealth}, \textit{Smart Cities}, etc.).

About the \textit{eHealth} scenario, all the sensors data available in the  \textit{tracker} environment (\textit{temperature}, \textit{humidity}, \textit{smog}, \textit{light level}, \textit{position}, \textit{altitude}, etc.) can be combined to those provided by a series of wearable sensors placed on the \textit{entity} (e.g., \textit{heart rate}, \textit{pressure}, etc.).
This configuration allows us to trace, in an exhaustive manner, the health status of an \textit{entity}, highlighting hidden \textit{person-environment} interactions, otherwise not obvious.

In other words, the data-flow existing between \textit{trackers} and \textit{entities} enrich the information provided by the individual sensors placed on an \textit{entity} body, since the \textit{IoE} environment allows us to add them the information related to all the sensors placed on the near involved \textit{tracker} devices.
This data-shared modality provides targeted (and more accurate) measurements and/or alerts, since it allows the system to have an overview of the real health-status of an \textit{entity}, with regards to a specific \textit{location} and with regard to some near \textit{entities}.

Similar interactions between \textit{entities} and \textit{trackers} can be also exploited in the \textit{Smart Cities} context, giving rise to a number of interesting applications.
Considering that the \textit{trackers} can be devices that operate, specifically, in such a context, their sensors data can be integrated to those related to a group of \textit{entities} in order to create functionalities aimed to specific groups of users.

This is an approach that leads towards two interesting advantages: it is able to uncover implicit characteristics of the involved \textit{entities} by following non canonical criteria~\cite{chong2018exploiting,saia2015latent}; each group of \textit{entities} can be anonymously characterized on the basis of the sensors data of the \textit{entities} that belong to it.

In light of the previous observations, we can consider the \textit{security} as one of the possible application scenarios of the \textit{IoE} paradigm proposed in this paper, which main scientific contributions have been summarized in the following:

\begin{enumerate}[(i)]
\item introduction of the novel concept of \textit{entities} and \textit{trackers}, able to exchange roles, which operates within a specific \textit{wireless-based} environment;
\item definition of interaction models between \textit{entities} and \textit{trackers}, and \textit{trackers} and \textit{blockchain-based distributed ledgers}, in terms of unique identification of the involved devices and communication techniques/protocols;
\item formalization of the \textit{entity}-to-\textit{tracker} and \textit{tracker}-to-\textit{blockchain-based distributed ledger} communication protocol data structures;
\item definition of criteria able to trace an \textit{entity} by exploiting the previous \textit{blockchain-based distributed ledgers} registrations, on the basis of a series of, directly or indirectly, strategies.
\end{enumerate}

The paper is organized into the following sections: Section~\ref{BackgroundRelatedWorkSection} provides an overview about the background and related work; Section~\ref{FormalNotationSection} reports the adopted formal notation; Section~\ref{ApproachFormulationSection} describes the implementation of the proposed \textit{IoE} paradigm; Section~\ref{FutureDirectionsSection} discusses about some future directions related to \textit{IoE}; Section~\ref{ConclusionSection} closes the paper with some concluding remarks.

\section{Background and Related Work}
\label{BackgroundRelatedWorkSection}
This section aims to introduce the most important concepts related to the context taken into account in this paper, starting by offering an overview on the \textit{Mobile Network} and \textit{Internet of Thing} concepts, continuing by introducing the \textit{blockchain-based} applications together with other concepts that revolves around them, and concluding with some consideration about the security aspects related to the aforementioned scenarios.

\subsection{Mobile Network}
A \textit{mobile} (or \textit{cellular}) network is a \textit{wireless-based} network geographically distributed in a number of areas defined \textit{cells}\cite{rappaport1996wireless,kaufman2008cellular}.
This mechanism based on \textit{cells} divides the mobile network area into many of overlapping geographic areas.

\begin{figure}[!]
\centering
\scriptsize
\includegraphics[width=0.6\textwidth]{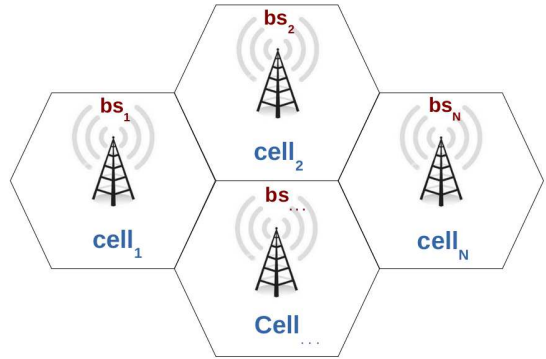}
\caption{$Mobile\ Network\ Structure$}
\label{MobileNetworkStructureGraph}
\end{figure}

It can be imagined as a mesh of hexagonal \textit{cells}, where each \textit{cell} has a base-station ($bs$) at its center, as shown in Figure~\ref{MobileNetworkStructureGraph}.
A slight overlapping between neighbor cells offers to the mobile devices a continue radio coverage, since in this way they are covered by at least  one base-station.
Such a base-station that serves a cell works as a hub, since the radio signal transmitted by a mobile device is retransmitted from the base-station to an other mobile device, transmitting and receiving by adopting different frequencies in order to avoid interferences.
In addition, the base-stations are connected through a central switching service that allows them to track the mobile device calls, transferring these from a base-station to another one, when a mobile device moves between cells.

The most important characteristics of the current mobile network that can be profitable exploited in the proposed \textit{IoE} paradigm are the wide coverage (that offer us a stimulating initial environment) and the high bandwidth (that allows us to quickly transfer the data between \textit{entities} and  \textit{trackers} and between \textit{trackers} and \textit{distributed ledgers}).

\subsection{Internet of Things}
In recent years we have seen how \textit{Internet} has given life to a new revolution that involves billions of devices.
These are characterized by both a low-cost and a capability to communicate in wireless way through \textit{Internet} and they are the main actors of this revolution named \textit{Internet of Things }(\textit{IoT}).
Into the \textit{IoT} environment operates heterogeneous devices, such as \textit{computers}, \textit{smart-phones}, \textit{wearable devices}, \textit{IP cameras}, \textit{RFID devices}, as well as a large number of actuators and sensors based on low-cost hardware, which represent the backbone of the \textit{IoT} environment.

This gives life to a kind of ecosystem founded on the communication paradigm, considering that each device can communicate with other devices and all the devices can communicate with each other without any geographic limitation, thanks to \textit{Internet}.
Another important \textit{IoT} characteristic is that each connected device is uniquely identified.

\begin{figure}[!]
\centering
\scriptsize
\includegraphics[width=0.8\textwidth]{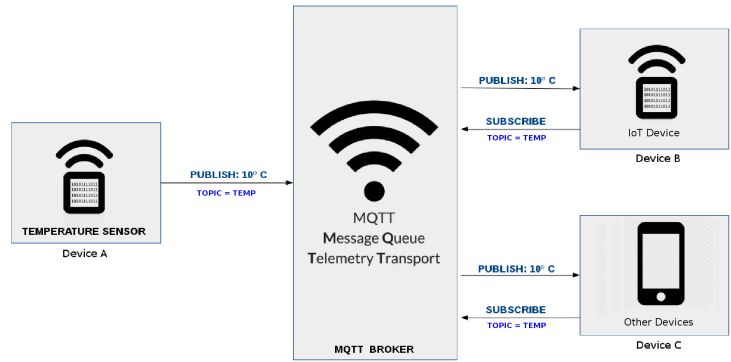}
\caption{$IoT\ Communication\ Paradigm$}
\label{IotArchitectureGraph}
\end{figure}

Premising that an \textit{IoT} device is potentially able to communicate directly with another one, a common \textit{IoT} communication paradigm is that exemplified in Figure~\ref{IotArchitectureGraph}: each device communicate to the other ones through two basic activities, \textit{publishing} and \textit{subscription}; it uses a protocol in order to \textit{publish} data on a server defined \textit{Broker} conventionally (in the example of Figure~\ref{IotArchitectureGraph}, it uses one of the most common \textit{IoT} protocols, \textit{MQTT}\footnote{Message Queue Telemetry Transport}); other devices can \textit{subscribe} the published data by selecting the \textit{topic} where it has been stored; the \textit{topic} represents the channel that allows a selective intercommunication between \textit{IoT} devices.

\subsubsection{Internet of Everything}
The growth of the \textit{Internet of Things} model and, more generally, the growth of the \textit{wireless-based} technologies, has contributed to the definition of a model called \textit{Internet of Everything}~\cite{yang2017internet,hussain2017internet}.
Such a model is characterized by the integration of \textit{data}, \textit{processes}, \textit{things}, and \textit{people}, combining several elements that in the past were separated from each other.
Some cases in point are the smart things such as \textit{smart-watches}, \textit{eHealth-devices}, and \textit{smart-vehicles}.

In other words, the \textit{Internet of Everything} model is used to refer to the intelligent interconnection of \textit{data}, \textit{processes}, \textit{things}, and \textit{people}, a scenario that involves billions of objects connected over public or private networks by using different protocols (\textit{standard} or \textit{proprietary}), which are able to detect the environment around them (\textit{sensors}) and/or able to interact with it (\textit{actuators}).

Summarizing, the \textit{Internet of Everything} model is different from the \textit{Internet of Things} one, since its paradigm is based on four elements (\textit{data}, \textit{processes}, \textit{things}, and \textit{people}), instead of one (\textit{things}).

\subsubsection{Identity of Things}
The \textit{Identity of Things} represents a concept mainly related to the \textit{Internet of Things} environment.
Basically, it refers to the need to assign an unique identifier to all objects that operates in such a environment, in order to allow their real-time interaction with people and other objects (\textit{things}).
A centered and quite recent example of the aforementioned scenario is that of the \textit{autonomous vehicles}~\cite{gerla2014internet}, where the concept of unique identification becomes day after day even more crucial~\cite{shao2007leader}.

The identifier can be created by using information that characterize, uniquely, the \textit{IoT} device such as, for instance, the manufacturer, the serial number, and so on.
Alternatively, the identifier can be assigned to the \textit{IoT} device by using a centralized or decentralized assignation remote service, manually or automatically.
Some possible approaches able to perform this operation are presented in Section~\ref{ElementsDefinitionSubsection}.

\subsection{Blockchain-based Applications}
A \textit{blockchain}, in the context of the cryptocurrency applications such as \textit{Bitcoin}~\cite{nakamoto2008bitcoin,danezis2015centrally} and \textit{Ethereum}~\cite{wood2014ethereum}, represents a shared and transparent \textit{distributed ledger}.
It allows the users to perform secure financial transaction by exploiting a cryptographic mechanism and it can be imagined as a ever-growing chain of blocks, where each block stores a sequence of transactions that are freely \textit{inspectable} by anyone but that are \textit{tampering-proof}.
Each of these blocks contains the cryptographic signature of the previous one and this mechanism does not allow anyone to \textit{alter} or \textit{remove} a previous block without the removal of all the blocks after it.

The \textit{blockchain} functionality can be exploited also in non-financial contexts, in all the cases where an application needs to ensure trust services. 
In other words, such a technology can be used as a platform to define the underlying trust level of an application.
The \textit{blockchain} ability to verify an identity through a reliable authentication process~\cite{pilkington201611} is indeed exploited in the context of heterogeneous environments, such us, for instance, those related to the \textit{eHealth}~\cite{kuo2017blockchain,castaldo2018blockchain}, \textit{smart cities}~\cite{biswas2016securing}, and \textit{IoT}~\cite{xu2018blockchain} applications.

Generalizing the concept, the \textit{blockchain} can be be profitably used in all the applications where there is the need to \textit{identify} an object (people, vehicles, documents, etc.) in a certain way.
For instance, it is used in~\cite{abbasi2017veidblock} to get a verifiable identity through a reliable authentication process, in~\cite{yuan2016towards} in order to introduce \textit{blockchain-based} intelligent transportation systems, in ~\cite{muftic2017blockchain}, where the \textit{blockchain} has been exploited to define a public identities ledger in the context of an identity management system, and in~\cite{ainsworth2016blockchain} in order to face the \textit{Value Added Tax} (\textit{VAT}) fraud problem.

\subsubsection{Double-spending Issue}
The \textit{double-spending} issue arises due to the absence of a central intermediary.
Explaining it in a few words, we suppose that \textit{Alice} has \textit{100-coins} and send all of them to \textit{Bob}: the \textit{double-spending} problem is related to the fact that \textit{Bob} can not know that \textit{Alice} had sent the same \textit{100-coins} to another person (e.g., \textit{Charlie}), because there is not a central intermediary (e.g., a bank) that verify such a transaction.

\begin{figure}[!]
\centering
\scriptsize
\includegraphics[width=0.7\textwidth]{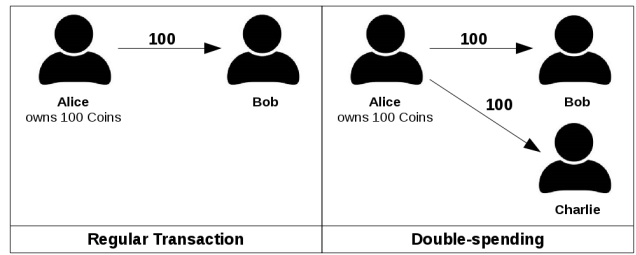}
\caption{$Double\mbox{-}spending\ Issue$}
\label{DoubleSpendingIssue}
\end{figure}

This problem, graphically summarized in Figure~\ref{DoubleSpendingIssue}, has been faced by adopting a distributed \textit{time-stamp} mechanism able to determine which transactions should be accepted and which should be rejected.
In the context of the \textit{Bitcoin} has been adopted a \textit{hash-chain} mechanism to perform this operation~\cite{kuo2017blockchain}.

Relating to the previous example, hypothesizing that the transaction from \textit{Alice} to \textit{Charlie} is stored in $block\mbox{-}1$ and the transaction from from \textit{Alice} to \textit{Bob} is instead stored  in $block\mbox{-}2$: through the \textit{hash-chain} mechanism each participant can verify that $block\mbox{-}1$ is older than $block\mbox{-}2$ by verifying the hashed \textit{blockchain}, avoid a \textit{double-spending} event by rejecting the transaction from \textit{Alice} to \textit{Bob}.

\subsubsection{Consensus Mechanism}
The \textit{consensus mechanism} stands at the base of the \textit{blockchain} paradigm, since it allows the system to append new blocks to the \textit{blockchain}.
In order to perform this operation, this mechanism exploits the so-called \textit{proof-of-work} (\textit{PoW})~\cite{jakobsson1999proofs}, a criterion based on the solution of a mathematical cryptological problem that involves as input the transactions stored into the block to add to the \textit{blockchain}. 

Literature defines as \textit{miners} the users that operate in order to solve this kind of problem.
When a \textit{miner} finds its solution, it is communicated to all the other users, who confirm its correctness and validate the new block, allowing the system to append it to the \textit{blockchain}.

The \textit{PoW} has been introduced during the \textit{Bitcoin}~\cite{nakamoto2008bitcoin} formalization and it assumes that each \textit{peer}\footnote{Equipotent node: in our case, it represents each \emph{blockchain} participant.} votes by using its \textit{computational power} by solving the mathematical cryptological problem and adding the current block to the \textit{blockchain}.
This mechanism, based on the users consensus, is aimed to protect the system against alterations and other fraudulent activities, since the \textit{PoW} activity (i.e., the solution of the mathematical cryptological problem) needs a very high computation load, which involves resources that are not normally available for a single user or for a small group of users.

It should be noted that the literature offers other consensus mechanisms to use instead of the \textit{PoW} one, such as, for instance, the so-called \textit{Proof of Stake} (\textit{PoS})~\cite{lin2017survey}.

\subsubsection{Distributed Ledger Technology}
As it emerges from the cited literature examples, the core of each application based on the \textit{blockchain} infrastructure is the \textit{Distributed Ledger Technology} (\textit{DLT}).
It is indeed clear how the identification process relies on the functionality offered by such a ledger, which protects the \textit{anonymity} of the \textit{entities}, assuring at the same time a certain identification. 

\begin{figure}[!]
\centering
\scriptsize
\includegraphics[width=1.0\textwidth]{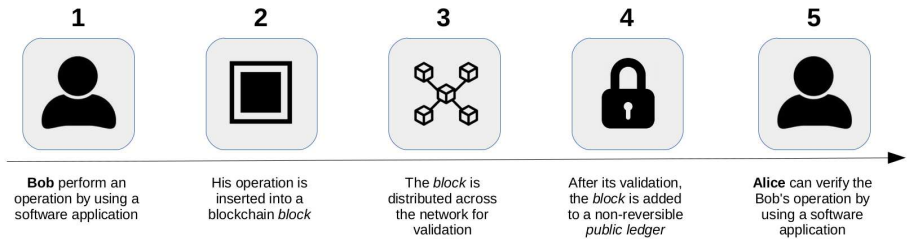}
\caption{$Blockchain\ Distributed\ Public\ Ledger$}
\label{DisLedgerArchitectureGraph}
\end{figure}

The process of \textit{insertion} and \textit{validation} of an operations (e.g., a financial transaction), carried out by using a \textit{distributed public ledger} based on the \textit{blockchain}, has been exemplified in Figure~\ref{DisLedgerArchitectureGraph}.

Through a \textit{blockchain} is possible to implement two different type of ledger: \textit{unpermissioned ledger} and \textit{permissioned ledger} (also known as \textit{private blockchain}).
Well-known example of \textit{unpermissioned ledger} are the \textit{Bitcoin} and \textit{Ethereum} environments, which have been designed to be \textit{open} and \textit{uncontrolled}. 
In more detail, they do not have single owners and this means that they allow anyone to add data to the ledger and any user that uses the ledger has identical copies of it. 
Users maintain the ledger integrity by reaching a consensus about its state. 

The \textit{unpermissioned ledger} are different from the \textit{permissioned} ones, which are ledgers where the users need a permission to get the access to them. 
This means that, when a new record is added to the ledger, its integrity is checked by following a restricted consensus process. 
The \textit{blockchain-based} \textit{permissioned ledgers} add a security level, since the consensus process generates a digital signature.

In the aforementioned context the term \textit{unpermissioned} is then a synonym of the term \textit{uncontrollable}, and for this reason most of the implementation adopt the \textit{unpermissioned} paradigm, since it is the only one able to provide the \textit{decentralization} that stand at the base of the \textit{blockchain} philosophy.

The working model adopted by any \textit{blockchain-based} ledger is based on the complete storage of all the information since its creation.
Such a model produces a constant and continue increasing of its size and this generates a crucial \textit{scalability issue} that must be effectively faced in the future~\cite{croman2016scaling}.
By way of example, at the beginning of \textit{2018} the size of \textit{Bitcoin} ledger has been evaluated about \textit{145-gigabyte} and that of \textit{Ethereum} ledger about \textit{40-gigabyte}~\cite{benvcic2018distributed}.

\subsubsection{Decentralized Storage Network}
The \textit{blockchain-based} technology also given rise to a new \textit{decentralized model} on which the \textit{Decentralized Storage Networks} (\textit{DSN}) are based.
By adopting this model, instead of using many servers (i.e., a \textit{server farm}), as it happens by adopting a canonical \textit{centralized storage model}, every network user (\textit{node}) stores part of network data.
Each user has an incentive to be part of the system and to keep the data available, for reasons similar to those that regulate the \textit{BitTorrent} file distribution system, a protocol that adopts a decentralized model that exploits the capability of the participants to network \textit{peer-to-peer} among themselves. 

Some examples of \textit{blockchain-based} decentralized storage approaches are \textit{Filecoin}\footnote{https://filecoin.io/}, \textit{SAFE Network}\footnote{https://safenetwork.org/}, \textit{Swarm}\footnote{https://github.com/ethereum/go-ethereum/tree/master/swarm}, \textit{Storj}\footnote{https://storj.io/}, and \textit{Sia}\footnote{http://sia.tech/}.

Considering that the traditional distributed models adopted for the cloud storage services, such as, for instance, those offered by \textit{Amazon's cloud storage}\footnote{https://aws.amazon.com/}, represent a market that generates profits in the billions of dollars: the growth of the decentralized storage model is cutting out part of this total market.
In addition to offer to the users a cheaper way for data storage, such a decentralized model also contributes to increase the availability of storage space.

\subsubsection{Blockchain and IoT Integration}
Scenarios characterized by the integration of \textit{blockchain-based} infrastructures with \textit{IoT} devices have been discussed in literature, such as in~\cite{reyna2018blockchain}, where the authors have been identified the following operative modalities:

\begin{itemize}
\item \textit{IoT-IoT}: it is characterized by a low-latency and an high-level of security, since the involved \textit{IoT} devices operates between them for most of the time, by exploiting the canonic protocols and by limiting the \textit{blockchain} use for storing only few information;\\
\item \textit{IoT-Blockchain}: by following this strategy, all the \textit{IoT} information are stored on the \textit{blockchain}, assuring their immutability and traceability, but increasing the bandwidth consumption and the latency-time;\\
\item \textit{Hybrid Paradigms}: this last strategy combines the aforementioned ones, performing part of the activities directly between the \textit{IoT} devices, limiting to the data storage activity the interaction with the \textit{blockchain}.
\end{itemize}

For the needs of the proposed \textit{IoE} paradigm, the second and third strategies (i.e., \textit{IoT-Blockchain} and \textit{Hybrid}) are the most suitable, although by adopting optimized criteria, the best strategy results the \textit{Hybrid} one, since it is the only one that allows us to balance the advantages offered by the \textit{IoT-IoT} and \textit{IoT-Blockchain} strategies.

\subsection{Security Aspects}
Some considerations should also be made about the security scenario related to the \textit{wireless-based} technologies, since such a technological evolution did not keep up with the security one.
It means that the big opportunities offered by the new technologies have been jeopardized by a series of problems that affect the security in a broad sense.

Some cases in point are the frauds related to the E-commerce infrastructure, which we have been dealt with in~\cite{saia2015multiple,saia2015proactive,DBLP:conf/nss/Saia17,DBLP:conf/secrypt/SaiaC17,DBLP:conf/iotbd/SaiaC17,DBLP:conf/iotbd/Saia18}, where retroactive, proactive, transformed-domain-based, and multidimensional  approaches have been experimented in order to face such problems, as well as the ever-increasing number of identity theft~\cite{bilge2009all,chou2004client} or, even more simply, the countless frauds made by exploiting the people's trust~\cite{jagatic2007social,arachchilage2016phishing}, often by recurring to \textit{social engineering} techniques~\cite{mouton2016social}.

Also in the \textit{mobile network} context we can observe similar problems, because the smart devices that operate in this environment inherit the security risks that characterize the \textit{Internet-based} devices (e.g., \textit{desktop computer}, \textit{laptop}, and so on), such as the aforementioned ones.
In addition, there are a series of more specific risks related to this context~\cite{liu2009cellular} such as, for instance, those related to the \textit{bot-net-based} attacks~\cite{traynor2009mitigating}, or those that jeopardize the user privacy~\cite{firoozjaei2017privacy}.

Even with regard to the \textit{blockchain-based} technologies (e.g., those related to the \textit{cryptocurrency}), their potential advantages have been flanked by a series of security issues related to the criminal efforts, which are aimed to exploit those new technologies, fraudulently.
In this specific case, the security issues have been boosted by the fact that such criminal activities can not be easily detected by surveillance authorities~\cite{Moore13ijcip}.

An example of security issue is related to the \textit{blockchain} consensus mechanism needed to add a new block, which involves many people called \textit{miners} that spend computation time (\textit{GPU}/\textit{CPU} type) to solve a kind of mathematical problem (\textit{hash-checking}).
A group of people can operate jointly as a \textit{mining-pools} in order to mining many blocks and this can leads towards the \textit{blockchain} control, if the achieved computing \textit{power} is at least the $51\%$ of the total~\cite{courtois2014subversive,eyal2018majority}.
This type of attacks are known in literature as \textit{Majority Attack} and they have been also theorized in the famous \textit{Satoshi Nakamoto} \textit{Bitcoin} white-paper~\cite{nakamoto2008bitcoin}. 

It should be observed how the \textit{PoW} mechanism that stands at the base of the \textit{blockchain} paradigm should not be too hard to solve, in order to avoid a very long block generation time that would bring toward the total block of all the transactions.
However, such a problem can not be overly simple to solve, because in this case the system would be vulnerable to many types of attacks such as, for instance, the \textit{Denial of Service} (\textit{DoS}) one~\cite{needham1993denial,vasek2014empirical}.

Other cases in point about the security issues in this context are the vulnerabilities that affect the \textit{Ethereum} \textit{smart contracts}~\cite{atzei2017survey} and the fraudulent games implemented through the \textit{blockchain} platform, such as those based on the well-known \textit{Ponzi schemes}~\cite{Artzrouni09,artzrouni2009mathematics}.
They have been introduced on the web many years ago~\cite{Moore12fc,lewis2012new} and recently re-proposed on \textit{Bitcoin}~\cite{Vasek15fc} and \textit{Ethereum}~\cite{bartoletti2017dissecting}.

\section{Formal Notation}
\label{FormalNotationSection}
Considering that we use the term \textit{entity} to indicate a device designed to operate in a \textit{IoE} environment, associated to a person or thing, and that we use the term \textit{tracker} to indicate a generic (new or already existing) device that operates in a \textit{wireless-based} environment, which is aimed to interact with the \textit{entities}, we introduce the following formal notation:

\begin{enumerate}[(i)]
\item we denote as $E=\{e_1,e_2,\ldots,e_M\}$ a set of \textit{entities}, and we use $E(e)$ to indicate such information related to an \textit{entity} $e$;;
\item we denote as $E_{\tau}=\{e_1,e_2,\ldots,e_N\}$ the \textit{entities} in $E$ detected by a \textit{tracker} device within $\tau$ seconds after the detection of an \textit{entity} (then $E_{\tau}\subseteq E$), and we use $E_{\tau}(e)$ to indicate such information related to an \textit{entity} $e$;
\item we denote as $L=\{l_1,l_2,\ldots,l_O\}$ a set of geographic locations, with $l=\{latitude,longitude\}$, and we use $l(e)$ to indicate such information related to an \textit{entity} $e$, when it is detected by a \textit{tracker} device;
\item we denote as $T=\{t_1,t_2,\ldots,t_P\}$ a set of \textit{time-stamps}, with $t=\{yyyy\mbox{-}mm\mbox{-}dd\mbox{-}hh\mbox{-}mm\mbox{-}ss\}$, and we use $t(e)$ to indicate the \textit{time-stamp} related to the detection of an \textit{entity} $e$ by a \textit{tracker} device;
\item we denote as $I=\{i_1,i_2,\ldots,i_Q\}$ a set of ($GUIDs$)\footnote{\textit{Globally Unique IDentifiers}, whose structure is formally defined in the \textit{RFC-4122}, which is explained in Section~\ref{ElementsDefinitionSubsection}.}, using the notation $i(e)$ to indicate the $GUID$ associated to an \textit{entity} $e$, as well as the notation $i(tracker)$ to indicate the $GUID$  associated to a \textit{tracker} device;
\item we denote as $P=\{p_1,p_2,\ldots,p_W\}$ a \textit{payload}, with $p=\{key,value\}$, and we use $P(e)$ to indicate a \textit{payload} related to an \textit{entity} $e$;
\item we denote as $R=\{r_1,r_2,\ldots,r_Y\}$ a set of registration made on a \textit{blockchain-based} distribute ledger, with $r=\{i(e),E_{\tau}(e),l(e),t(e),P(e)\}$, and we use $r(e)$ and $R(e)$ to indicate, respectively, a registration related to an \textit{entity} $e$ and all the registrations related to that \textit{entity}.
\end{enumerate}

\section{Approach Formulation}
\label{ApproachFormulationSection}
This section describes the implementation of the proposed \textit{IoE} paradigm, which has been divided in the following steps:

\begin{enumerate}[(i)]
\item \textbf{Elements Definition}: it introduces the concept of \textit{entity} and \textit{tracker} in the \textit{IoE} environment, as well as the method to use in order to assign them a \textit{Globally Unique Identifier}, outlining some possible operative scenarios;
\item \textbf{Elements Detection}: the detection process of an \textit{entity} device is here described, from the \textit{detection-time} by a \textit{tracker} device to the \textit{recording-time} of the collected data on a \textit{blockchain-based distributed ledger}, focusing on the characteristics of the state-of-the-art wireless technologies able to perform these activities;
\item \textbf{Elements Communication}: it formalizes the data structures and the software procedures able to merge the information related to the involved \textit{entity} and \textit{tracker} devices, generating the \textit{data-structure} that represent the information to store on the \textit{blockchain-based distributed ledger};
\item \textbf{Elements Localization}: extensively, it describes the activities made in order to trace an \textit{entity}, introducing some baseline strategies and a series of localization rules aimed to exploit the available information on the \textit{blockchain}, directly or indirectly.
\end{enumerate}

\begin{figure}[!]
\centering
\scriptsize
\includegraphics[width=0.8\textwidth]{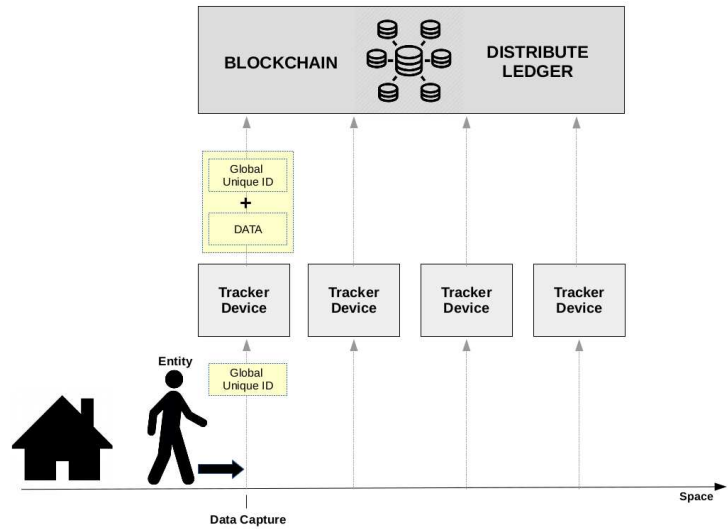}
\caption{$IoE\ Working\ Model$}
\label{IoeArchitectureGraph}
\end{figure}

\subsection{Elements Definition}
\label{ElementsDefinitionSubsection}
The concept of \textit{entity} is usually related to a person, but it could be also extended to a large number of objects such as, for instance, \textit{vehicles} or \textit{goods}, and each \textit{entity} $e$ is always associated to a \textit{Globally Unique Identifier} (\textit{GUID}).

The concept of \textit{tracker} is instead related to a generic device able to detect the \textit{entity} devices, capturing their \textit{GUID}s and sensors data, and performing a registration into a \textit{blockchain-based distributed ledger}.
Such a registration (i.e., the set $r$) is defined by joining \textit{entity} and \textit{tracker} data, according to the formal notation defined in Section~\ref{FormalNotationSection}.

\scriptsize
\begin{lstlisting}[style = mystyle,label={GuidListing},caption={$Globally\ Unique\ Identifier\ Data\ Structure$}]
GUID = 	time-low "-" time-mid "-"
		time-high-and-version "-"
		clock-seq-and-reserved
		clock-seq-low "-" node
time-low = 4hexOctet
time-mid = 2hexOctet
time-high-and-version = 2hexOctet
clock-seq-and-reserved = hexOctet
clock-seq-low = hexOctet
node = 6hexOctet
hexOctet = hexDigit hexDigit
hexDigit = "0" / "1" / "2" / "3" / "4" / "5" / "6" / "7" / "8" / "9" /
		   "a" / "b" / "c" / "d" / "e" / "f" /
		   "A" / "B" / "C" / "D" / "E" / "F"
\end{lstlisting}
\normalsize

The unique identifier of the \textit{tracker} devices could be already available (e.g., \textit{MAC-address}, \textit{IP-address}, etc.), while that of the new \textit{entity} devices placed in the \textit{IoE} environment needs to be defined and assigned.
Its generation can be made in several ways~\cite{jones2012creating,watson1981identifiers}, but the two most common methods are: (\textit{i}) on the basis of a \textit{serial numbers} created by following an incremental or sequential criterion; (\textit{ii}) on the basis of a \textit{random numbers} generated by using a range of numbers enough larger to classify the expected number of objects.
In the proposed approach, we perform this operation by using one of the most effective methods: the  \textit{Globally Unique Identifier}.

\textit{}\\
\textbf{Globally Unique Identifier}: The \textit{Globally Unique Identifier} (\textit{GUID}), also known as \textit{Universally Unique Identifier} (\textit{UUID}), is a \textit{128-bit} integer number which is commonly used in order to identify resources uniquely~\cite{leach2005universally}.
When it needs, such a information can be combined with additional information (e.g., related to one or more resource characteristics) in order to identify the same resource in different contexts.
Listing~\ref{GuidListing} reports the formal definition of a \textit{GUID} string, and \textit{f81d4fae-7dec-11d0-a765-00a0c91e6bf6} represents an example of this one.
Several algorithms able to generate this information are described in~\cite{leach2005universally}.

Through the application of the \textit{birthday paradox}~\cite{hankerson2004cryptographic,mironov2005hash} we can obtain a mathematically demonstration of the \textit{GUID} robustness in terms of hash collision probability.
By following this mathematical approach, considering that a \textit{GUID} is a \textit{128-bit} long number, we can identify a million billion \textit{entities} before we have a one in a billion possibility (i.e., $10^{15}$) to get a collision, as shown in Equation~\ref{CollisionProbability}, which is based on the aforementioned \textit{birthday paradox}.

\begin{equation}
n \approx \sqrt{-2^{129} \cdot ln(1 - 10^{-9}}) \approx 1,000,000,000,000,000
\label{CollisionProbability}
\end{equation}

Some considerations can be made about the policies to adopt in order to assign the \textit{GUID} to each \textit{entity} device that operates into the \textit{IoE} environment, assuring that this information remains stable along the time.
This because the \textit{IoE} tracing mechanism is based on such information and a change of it (i.e., the device \textit{GUID}) during the life of an \textit{entity} device leads towards inconsistent data.

Some solutions involve or a centralized \textit{GUID} distribution, such as in~\cite{manku2003symphony}, offered as service to the users by following a free or paid modality, or an autonomous generation of this information made directly by the users~\cite{leach2005universally}.
It should be added that in order to distinguish the \textit{IoE} devices from the other classes of devices that operate in the \textit{wireless-based} environment, it is appropriate to reserve part of the \textit{GUID} information for this purpose.

\textit{}\\
\textbf{Operative Scenarios}: About the hardware to use in the \textit{IoE} environment in order to allow the \textit{entity} devices to interact with the \textit{tracker} ones, we can outline several scenarios:

\begin{enumerate}[(i)]
\item the \textit{entity} device is characterized by limited or absent hardware resources (e.g., \textit{CPU}, \textit{memory}, etc), then it performs the identification process by exploiting passive technologies such as, for instance, \textit{RFID}\footnote{Radio-Frequency IDentification.}.
In this first scenario, the \textit{tracker} device must be able to manage the identification process adopted by the \textit{entity};
\item the \textit{entity} device has hardware resources that allow it to adopt active technologies for the identification process (e.g., \textit{6LoWPAN} and  \textit{ZigBee}, both defined by the \textit{technical standard IEEE 802.15.4}).
This is the most common scenario, where the \textit{entity} device uses canonical wireless technologies and the \textit{tracker} device does not need any additional capability in order to interact with it;
\item the \textit{entity} device is able to perform processes that require considerable hardware/software resources.
Such a scenario allows us to move on the \textit{IoE}-side some processes usually performed in the \textit{tracker}-side and it also allows the \textit{IoE} device to handle complex processes related to its sensors. 
\end{enumerate}

The scenario taken into consideration in this paper is the second one, where the \textit{IoE} device is characterized by enough hardware/software resources that allow it to use active technologies for its identification, because it allows us to implement the \textit{IoE} immediately and in a transparent way, postponing the other scenarios to possible future implementations.

\subsection{Elements Detection}
\label{ElementsDetectionSubsection}
As shown in the high-level working model of Figure~\ref{IoeArchitectureGraph}, when an \textit{entity} $e$ enters within the coverage area of a \textit{tracker} device, such a device detects its identifier $i$ (i.e., the \textit{GUID}, as formalized in Section~\ref{FormalNotationSection}), and it creates and submits a registration $r$ on a \textit{blockchain-based distributed ledger}.

The detection time of an \textit{entity} $e$ is indicated in Figure~\ref{IoeArchitectureGraph} as \textit{data capture} and it coincides with the \textit{time-stamp} $t$, which represents the point in the space where the \textit{entity} is detected by a \textit{tracker} device and the $r$ information are submitted to the \textit{blockchain-based distributed ledger}.
   
All the above operation are managed by using specific data structures, whose possible implementation has been proposed in Section~\ref{ElementsCommunicationSubsection}.

\textit{}\\
\textbf{Wireless Technologies}: About the technology to use in order to broadcast the \textit{entity} \textit{GUID}, the literature offers several solutions in terms  of technologies and protocols able to perform this operation~\cite{al2017internet}.
Some examples of them are: \textit{Internet Protocol Version 6 over Low-Power Wireless Personal Area Networks} (\textit{6LoWPAN}), \textit{Bluetooth Low Energy} (\textit{BLE}), \textit{Z-Wave}, \textit{ZigBee}, \textit{Near Field Communication} (\textit{NFC}), \textit{Radio Frequency IDentification} (\textit{RFID}), \textit{SigFox}, and \textit{2G/3G}.
\textit{SigFox} and \textit{2G/3G} are classified as \textit{Low-Power Wide Area Network} (\textit{LPWAN}) protocols, while the other ones as \textit{Short-range Wireless} protocols.

Their characteristics have been summarized in Table~\ref{WirelessProtocolsTable}, where the reported ranges (i.e., \textit{frequency range} and \textit{operative range}) indicates only the lowest and the highest supported value (e.g., if the protocol supports $125KHz$, $13.56MHz$, and $860MHz$, we report $125KHz \div 860MHz$).

The choice of protocol should be made by taking into account the \textit{entity} type, since in case of a \textit{person} such a choice should be oriented toward protocols able to ensure a low-power consumption and a mid/short operative range, while in case of \textit{objects} (e.g., a vehicle) the choice could be instead oriented toward protocols characterized by a long operative range and a mid/high power consumption.

However, the above considerations are strongly related to the context of a custom \textit{IoE} device, since when it is a standard device such as, for instance, a \textit{smart-phone} or a \textit{tablet}, the choice of the wireless protocols is driven by those supported by the operating system (e.g., \textit{802.11 b/g/n}~\cite{uzcategui2009wave} and \textit{Bluetooth Low Energy} (\textit{BLE})~\cite{gomez2012overview} protocols).

\begin{table}[ht]
\caption{Wireless Technologies}
\label{WirelessProtocolsTable}
\centering
\tiny
\resizebox{12.0cm}{!} {
\begin{tabular}{l|l|l|l|l|l|l} 
\toprule
{Wireless} & {Frequency} & {Data} &{Operative} &{Power} &{Security} &{Literature} \\
{technology} & {range} & {rate} &{range} &{consumption} &{protocols} &{reference} \\
\toprule
\textbf{6LoWPAN} & {868MHz$\div$2.4GHz}  & {250KBps}           & {10$\div$100m} & {low} & {AES} & {\cite{mulligan20076lowpan}}\\ 
\textbf{BLE} 	 & {2.4GHz}  			 & {1MBps}             & {15$\div$30m}  & {low} & {E0, Stream, AES-128} & {\cite{gomez2012overview}}\\
\textbf{Z-Wave}  & {868MHz$\div$908MHz}  & {40KBps}            & {30$\div$100m} & {low}& {AES-128} & {\cite{kuzlu2015review}}\\
\textbf{ZigBee}  & {2.4GHz}  			 & {250KBps}  		   & {10$\div$100m} & {low}  & {AES} & {\cite{kinney2003zigbee}}\\
\textbf{NFC}  	 & {868MHz$\div$902MHz}  & {106$\div$424KBps}  & {0$\div$1m}    & {Ultra-low} & {RC4} & {\cite{cerruela2016state}}\\
\textbf{RFID}    & {125KHz$\div$928MHz}  & {4MBps}  		   & {0$\div$200m}  & {Ultra-low} & {RSA,AES} & {\cite{jia2012rfid}}\\
\midrule
\textbf{SigFox}  & {125KHz$\div$860MHz}  & {100$\div$600Bps}   & {10$\div$50Km} & {low}  & {no-specific} & {\cite{raza2017low}}\\
\textbf{2G/3G}   & {380MHz$\div$1.9GHz}  & {10MBps} 		   & {Several Kms}  & {High}  & {RC4} & {\cite{novo2015capillary}}\\
\bottomrule
\end{tabular}
}
\end{table}

\subsection{Elements Communication}
\label{ElementsCommunicationSubsection}
\begin{figure}[!]
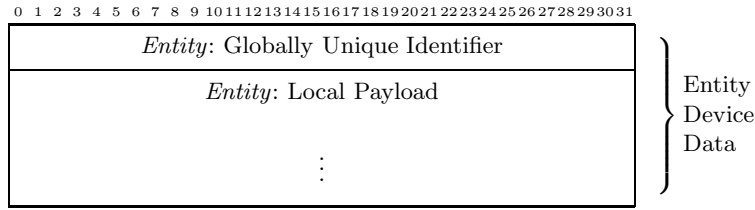

\centering
\begin{bytefield}[bitwidth=0.8em]{32}
        \bitheader{0-31} \\
        \begin{rightwordgroup}{{Entity} \\  Device\\ Data}
        \bitbox{32}{\textit{Entity}: Globally Unique Identifier} \\
        \wordbox[tlr]{1}{\textit{Entity}: Local Payload} \\
        \wordbox[blr]{2}{$\vdots$}   
        \end{rightwordgroup}         
\end{bytefield}
\caption{$\textit{Entity-side}\ Data\ Structure$}
\label{ProtocolPacketStructureIoe}
\end{figure}

The communication between an \textit{entity} $e$ and a \textit{tracker} device can be performed by adopting very simple data structures, whose possible formalization are proposed in Figure~\ref{ProtocolPacketStructureIoe} and Figure~\ref{ProtocolPacketStructureIot}.

They refer, respectively, to the data structure used to transmit data from an \textit{entity} device to a \textit{tracker} device (i.e., \textit{entity-side}) and to the data structure used to transmit the registration data from a \textit{tracker} device to the \textit{blockchain-based distributed ledger} (i.e., \textit{tracker-side}).

About the \textit{Entity-side} data structure, the \textit{GUID} information, which is \textit{128}-bit long, is stored by using \textit{5} groups of hexadecimal digits, with the following size: \textit{8} hexadecimal digits, \textit{4} hexadecimal digits, \textit{4} hexadecimal digits, \textit{4} hexadecimal digits, and \textit{12} hexadecimal digits. 

The registration data $r$ are defined by merging a series of identification data (\textit{Tracker Primary Data}) with the sensors data related both to the \textit{entity} and \textit{tracker} devices activity (\textit{Tracker Payload Data})
In some contexts, the \textit{Payload Data} could be partially (only the \textit{entity} or \textit{tracker} sensors data) or completely absent (no sensors data) and, in this cases, the \textit{entity} information will be the \textit{GUID}, the \textit{location}, and the \textit{time-stamp}.

About the hardware/software process performed in the \textit{entity-side}, it is limited to broadcast its data (\textit{GUID} and \textit{local payload}) at regular time intervals, by using the wireless functionality.
About the \textit{tracker-side} hardware/software process, when there are not active other priority tasks, the \textit{tracker} device operates a listening activity aimed to detect \textit{entities} in its wireless coverage area, sending the collected \textit{entity} and \textit{tracker} data to the \textit{blockchain-based distributed ledger}.

It should be observed that in the data structures we classified the \textit{payload} on the basis of the data which it refers, using the term \textit{local} to indicate that generated by the \textit{entity} device and \textit{global} to indicate that generated by the \textit{tracker} device, which also include the \textit{local payload}.

\begin{figure}[!]
\centering
\begin{bytefield}[bitwidth=0.8em]{32}
        \bitheader{0-31} \\
        \begin{rightwordgroup}{\textit{Tracker} \\ Primary \\  Data}
        \bitbox{32}{\textit{Entity}: Globally Unique Identifier}\\
        \wordbox[tlr]{1}{\textit{Entity}: Neighbor Entities List} \\
        \wordbox[blr]{2}{$\vdots$} \\
        \bitbox{16}{\textit{Tracker}: Latitude} & \bitbox{16}{\textit{Tracker}: Longitude}\\
        \wordbox[tlr]{1}{\textit{Tracker}: Timestamp}
        \end{rightwordgroup} \\
        \begin{rightwordgroup}{\textit{Tracker} \\ Payload \\ Data}
        \wordbox[tlr]{1}{\textit{Entity + Tracker}: Global Payload} \\
        \wordbox[blr]{2}{$\vdots$}
        \end{rightwordgroup}
\end{bytefield}
\caption{$\textit{Tracker-side}\ Data\ tructure$}
\label{ProtocolPacketStructureIot}
\end{figure}

The \textit{data anonymity} and \textit{data immutability} offered by a \textit{blockchain-based distributed ledger}, joined with the low-cost of the devices needed for the data transmission and with the wireless coverage offered by the ever increasing number of \textit{wireless-based} devices, given life to a powerful environment on which is based the proposed \textit{IoE} paradigm.
 
The data that we need to store on the \textit{blockchain-based distributed ledger} is that described in Section~\ref{FormalNotationSection}: the first field $i$ contains the \textit{Globally Unique Identifier} of the \textit{IoE} \textit{entity}; the field $E_{\tau}$ contains, when it is applicable, a list of \textit{Globally Unique Identifiers} related to the other \textit{entities} captured together with the \textit{entity} $e$ in a defined temporal frame $\tau$; the \textit{l} field contains the geographic position (i.e., \textit{latitude} and \textit{longitude}) of the \textit{tracker} device that detected the \textit{entity} $e$; the field \textit{t} reports when the \textit{data capture} event occurred, in the format \textit{yyyy-mm-dd-hh-mm-ss}; the last field \textit{P} contains a series of values in the format \textit{key,value} which refer to the sensors data of the \textit{entity} device (\textit{local payload}) and to the sensors data of the \textit{tracker} device (\textit{global payload}).

\textit{}\\
\textbf{Software Procedures}:
The software to use in order to perform the \textit{entity-tracker} and \textit{tracker-ledger} communications can be an update, in case of \textit{IoE} and \textit{custom devices}, or an application (\textit{app}), in most of the other cases (i.e., \textit{smart-phones}, \textit{tablets}, and similar devices).
It has to fulfill the \textit{IoE} paradigm needs, from the \textit{entity-detection} to the \textit{data-registration}, by performing the following operations:

\begin{enumerate}
\item \emph{entity-side}: it provides to broadcast the device \textit{GUID} along with the \textit{payload} (i.e., local sensors data), by using the built-in wireless device functionality;  
\item \textit{tracker-side}: it performs a listening activity aimed to detect and recognize (distinguishing them from the other devices through the mechanism adopted in the implementation phase, for instance, a specific \textit{GUID} preamble) \textit{entities} within its wireless coverage area;
\item \textit{tracker-side}: it appends the \textit{tracker} device data (i.e., \textit{primary} and \textit{payload} data) with the data transmitted by the \textit{entity} device (i.e., \textit{GUID} and \textit{payload}), building a data packet suitable for a registration on the \textit{blockchain-based distributed ledger};
\item \textit{tracker-side}: it submits the defined data packet on the \textit{blockchain-based distributed ledger}, in order to perform an immutable registration of the \textit{entity} device activity;
\item \textit{tracker-side}: it waits to receive from the \textit{blockchain-based distributed ledger} the registration acknowledge of the submitted packet, otherwise it repeats the submission.
\end{enumerate}

A series of custom \textit{data-dashboards}\footnote{A management tool able to display, track and analyze a series of information.} can be also designed in order to manage all the processes involved in the \textit{IoE} paradigm, first of all, that related to the constant tracking of the \textit{entities}.

\begin{algorithm}
\centering
\scriptsize
\caption{\small Blockchain-based distributed ledger data gathering}
\label{LedgerQueryAlgorithm}
\begin{algorithmic}[1]
\Require $e$=Entity, $R$=Blockchain-based distributed ledger registrations
\Ensure $\hat{R}$=Registrations related to entity $e$
\Procedure{getEntityRegistrations}{$e$, $R$}
	\For {\textbf{each} $r$ \textbf{in} $R$}	
		\State $i \leftarrow getEntityGUID(r)$
		\If{$i(e) == \hat{e}$}  
			\State $\hat{R} \leftarrow r(e)$
		\EndIf		
	\EndFor			
\State \textbf{return} $\hat{R}$	
\EndProcedure
\end{algorithmic}
\end{algorithm}

\subsection{Elements Localization}
\label{ElementsLocalizationSubsection}
When we need to investigate about an \textit{entity} $e$, first we get all needed data related to it by performing a \textit{data gathering} process, such as that reported in Algorithm~\ref{LedgerQueryAlgorithm}, then we can manage such data through different strategies, such as the baseline ones described below:

\begin{enumerate}
\item \textbf{Direct Tracing}: by following this strategy, the movements of an \textit{entity} $e$, from its first introduction in the \textit{IoE} environment, are traced by using the information $l(e)$ and $t(e)$ in $r(e), \forall r(e)\in R(e)$, according to the formalization given in Section~\ref{FormalNotationSection}.

This process is shown in Figure~\ref{IoeDirectTracingGraph}, which refers to six detection points $l$ of an \textit{entity} $e$, chronologically numbered by using the \textit{time-stamp} information $t$.
In more detail, we first query the \textit{blockchain-based distribute ledger} in order to extract all the registrations $R(e)$, then we number each location $l(e)\in r(e), \forall r(e)\in R(e)$  (i.e., \emph{latitude} an \textit{longitude}) along the chronological sequence given by the \textit{time-stamp} information $t(e)\in r(e)$.

More formally, given a series of \textit{entity} locations $l(e)\in L$, we introduce a \textit{Trace Location Set} $\omega=\{l_1,l_2,\ldots,l_Z\}$ aimed to store, in the chronologically order determined by the \textit{time-stamp} information $t(e)\in T$, all the locations $l(e)\in L$, as formalized in Equation~\ref{criterion0}.
\begin{equation}
\begin{array}{ll}
\omega \leftarrow \; l(e)\;\;\vert\;\; \forall\;\; l(e)\;\; \textbf{in}\;\; L\;\;\\
\textbf{with}\;\; l_1 < l_2 < \ldots < l_Z\;\; \wedge \;\; l\in \omega
\end{array}
\label{criterion0}
\end{equation}

It should be noted that the localization resolution is directly related to the \textit{tracker} device that has detected the \textit{entity}.
We can obtain a \textit{high-resolution} localization when the \textit{tracker} device runs a localization service (e.g., \textit{GPS}) and then its location is near that of the detected \textit{entity}.
We instead obtain a \textit{low-resolution} localization when the localization data are related to another device, as happens when the \textit{tracker} device operates in the mobile network but without any active localization service, since in this case the location could refer to the mobile network \textit{cell}.

This is represented in Figure~\ref{IoeDirectTracingGraph} and Figure~\ref{IoeIndirectTracingGraph}: the \textit{high-resolution} localization coincides with the \textit{entity} \textit{map-point}, while the \textit{low-resolution} localization can be considered any \textit{map-points} within the \textit{grid-square} where the \textit{entity} is placed, which represents the mobile network \textit{cell}.
\begin{figure}[!]
\centering
\scriptsize
\includegraphics[width=0.8\textwidth]{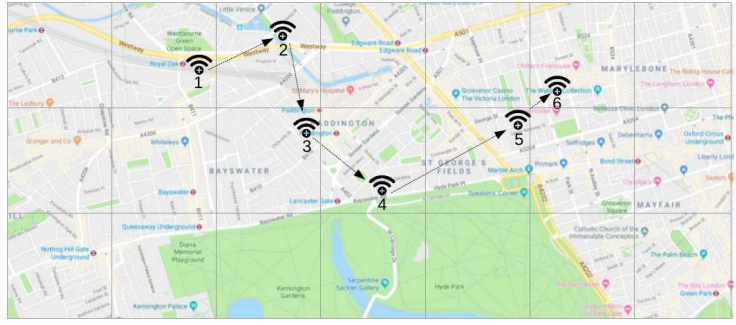}
\caption{$IoE\ Direct\ Tracing$}
\label{IoeDirectTracingGraph}
\end{figure}
\textit{}\\
\item \textbf{Interpolate Tracing}: in this strategy we take into account the information $l(e)$, $t(e)$, and $E_{\tau}(e)$ in $r(e), \forall r(e)\in R(e)$.
The $E_{\tau}(e)$ information contains, when it is applicable, the other \textit{entities} detected by the \textit{tracker} device within $\tau$ seconds from the $e$ (the \textit{entity} under analysis) detection, as described in Section~\ref{FormalNotationSection}.

We exploit the new information in order to reconstruct the \textit{entity} movements by interpolating the $l(e)\in r(e), \forall r(e)\in R(e)$ data with the same data of the \textit{entities} in $E_{\tau}(e)$ (neighbor \textit{entities}). 
This process is graphically shown in Figure~\ref{IoeIndirectTracingGraph}, where \textit{+} denotes the \textit{entity} under analysis and \textit{N} a neighbor \textit{entity} in $E_{\tau}(e)$.

In the example of \textit{interpolate tracing} shown in figure, we can observe how the first localization of the \textit{entity} \textit{+} includes a neighbor \textit{N} that we found another time in the third location of the location chronology of \textit{+}.
This represents a naive example of \textit{interpolate tracing}, based on the reasonable probability that such a configuration indicates that the neighbor \textit{entity} is somehow related to the main \textit{entity} under analysis, especially when this pattern repeats over time.
In other words, it is very likely that in the second localization of \textit{N}, the entity \textit{+} was also present, and that it has not been detected for some reasons such as, for instance, a temporary \textit{tracker} device overload, or because the \textit{entity} device was out of the \textit{tracker} wireless range.
This pattern, repeated over time, could underline interesting connections between \textit{entities}, as well as the last location of a missing \textit{entity}.

More formally, given the \textit{Trace Location Set} $\omega=\{l_1,l_2,\ldots,l_Z\}$ previously defined and given a series of \textit{entity} locations $L(e)=\{l_1,l_2, \ldots, l_O \}$, at each location $l\in L(e)$ (with $O\geq 3$) we extract from the set $E_{\tau}(e)$ a subset of valuable\footnote{Entities able to be exploited in the context of the \textit{Interpolate Tracing} strategy.} neighbor \textit{entities} by following the criterion in Equation~\ref{criterion1}.
\begin{equation}
\begin{array}{ll}
\omega \leftarrow \; l(e)\;\;\vert\;\; \textbf{if}\;\; e\;\; \textbf{in}\;\; E_{t}(l_{o-1})\;\; \wedge\;\; e\;\; \textbf{in}\;\; E_{t}(l_{o+1}), \;\; \forall\;\; e\;\; \textbf{in}\;\; E_{\tau}\\
\textbf{with}\;\; l_1 < l_2 < \ldots < l_Z\;\; \wedge \;\; l\in \omega
\end{array}
\label{criterion1}
\end{equation}
We can generalize the aforementioned criterion by varying the distance between the step $E_{\tau}(e)$ (i.e., where we extract the valuable neighbor \textit{entities} from $E(e)$) and the previous and next step that we take into account.
Denoting as $\alpha$ such a distance (i.e., the number of considered locations), we can re-formalize the former criterion as shown in Equation~\ref{criterion2}.
\begin{equation}
\begin{array}{ll}
\omega \leftarrow \; l(e)\;\;\vert\;\; \textbf{if}\;\; e\;\; \textbf{in}\;\; E_{t}(l_{o-\alpha})\;\; \wedge\;\; e\;\; \textbf{in}\;\; E_{t}(l_{o+\alpha})\\
\textbf{with}\;\; l_1 < l_2 < \ldots < l_Z\;\; \wedge \;\; l\in \omega
\end{array}
\label{criterion2}
\end{equation}

It should be underlined how during this activity we do not infringe the privacy of the involved \textit{neighbor entities}, since the \textit{entity} data are collected anonymously into the \textit{blockchain-based distributed ledger}.
\begin{figure}[!]
\centering
\scriptsize
\includegraphics[width=0.8\textwidth]{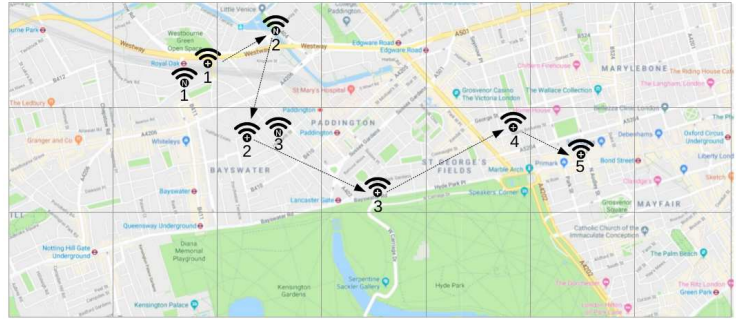}
\caption{$IoE\ Interpolate\ Tracing$}
\label{IoeIndirectTracingGraph}
\end{figure}

\textit{}\\
\item \textbf{Spread Tracing}: this last baseline criterion exploits all the neighbor \textit{entities} in $E_{\tau}$, with $e \neq \hat{e}$ and $\vert E_{\tau}\vert \geq 2$, where $\hat{e}$ denotes the \textit{entity} under analysis.

We add as valuable neighbor \textit{entities} of $\hat{e}$ all the \textit{entities} that in their locations in $L$ have $\hat{e}$ as neighbor \textit{entity}, as shown in  Equation~\ref{criterion3}.
\begin{equation}
\begin{array}{ll}
\omega \leftarrow \; l(e)\;\;\vert\;\; \textbf{if}\;\; \hat{e}\;\; \textbf{in}\;\; E_{\tau}(e), \;\; \forall\;\; l(e)\;\; \textbf{in}\;\; L\\
\textbf{with}\;\; l_1 < l_2 < \ldots < l_Z\;\; \wedge \;\; l\in \omega
\end{array}
\label{criterion3}
\end{equation}
The result can be expressed as the \textit{tracing matrix} $\Xi$ shown in Equation~\ref{EntityMatrixEquation}, where each row refers to a different valuable \textit{entity} $e$.
In other words, each matrix-row refers to a different valuable \textit{entity} $e$ and it reports the locations $l$ where the \textit{entity} $e$ has the \textit{entity} $\hat{e}$ as neighbor in $E_{\tau}(e)$.
\begin{equation}
\label{EntityMatrixEquation}
 \Xi(e) =
  \begin{bmatrix}
   l_1, & l_2, & \cdots, & l_O \\
   l_1, & l_2, & \cdots, & l_O \\
   \vdots & \vdots & \ddots & \vdots\\
   l_1, & l_2, & \cdots, & l_O \\
   \end{bmatrix}
\end{equation}
After ordering the matrix-row elements by location and after counting how many \textit{entities} $\hat{e}$ are involved in each matrix-column, we can evaluate the probability that the entity $e$ was in a specific position, although it has not been detected by a \textit{tracker} device.
This criterion is graphically shown in Figure~\ref{SpreadCriterion}, where the \textit{grid-size} (i.e., \textit{square-side}) represents a tolerance value, which we denoted as $\Delta$.
This meand that all the \textit{entity-detections} that occur into the same \textit{grid-square} refer to the same \textit{matrix-row-index} (i.e., Equation~\ref{EntityMatrixEquation}). 
\begin{figure}[!]
\centering
\scriptsize
\includegraphics[width=0.8\textwidth]{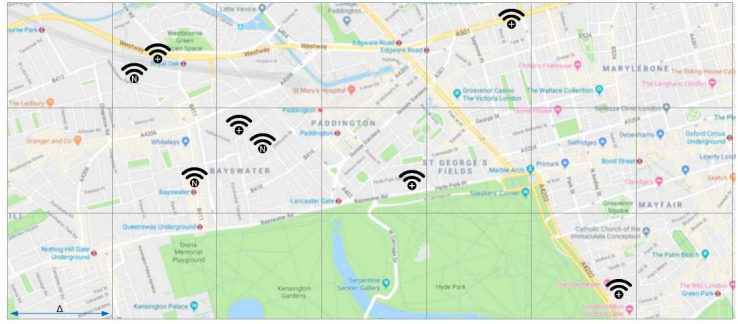}
\caption{$IoE\ Spread\ Tracing$}
\label{SpreadCriterion}
\end{figure}
\end{enumerate}

It should be noted how the grid of Figure~\ref{SpreadCriterion} represents a different information, with respect to that of Figure~\ref{IoeDirectTracingGraph} and Figure~\ref{IoeIndirectTracingGraph}, since in this case it does not represent the mobile network \textit{cells} but the tolerance value $\Delta$.

All the aforementioned criteria can be combined in order to define a more complex criterion based on different localization rules.
In addition, all criteria have been formalized by exploiting only few of the available information, which can be fully exploited in order to improve the localization strategies (e.g., by inferring further details from the sensors data).

\section{Future Directions}
\label{FutureDirectionsSection}
Considering that a complete and fully-functional implementation of the proposed \textit{IoE} paradigm is beyond the scope of this paper, which is mainly aimed to expose the theoretical concepts that revolve around our core idea, delineating several application scenarios, this section introduces some future directions, making also some general considerations about its potential spread.

\subsection{Secure Payload Storing}
\label{SecurePayloadStoringSubsection}
A future extension of the \textit{IoE} paradigm  could be designed in order to manage as \textit{payload} large and/or sensitive sensors data, by recurring both to external storage services and encryption protocols.
Such a problem arises with regard to the payload data generated by the \textit{tracker} device that detect an \textit{entity}, since such data could be refer to sensitive information generated by some classes of sensors such as, for instance, \textit{microphones} and \textit{video cameras}, instead than non-sensitive information generated by other classes of sensors (e.g., \textit{temperature sensors}, \textit{humidity sensors}, etc.).

A possible and effective solution able to face this problem is based on the \textit{asymmetric encryption} model~\cite{simmons1979symmetric}, which analogously to the canonical encryption mechanism adopted nowadays in a number of applications (e.g.,  \textit{SSH}, \textit{OpenPGP}, \textit{S/MIME}, etc.)\footnote{\textit{Secure Socket Shell}, \textit{Open Pretty Good Privacy}, \textit{Secure Multi-Purpose Internet Mail Extensions}}, is exploited in order to encrypt the data locally (when the \textit{tracker} functionalities allow us this operation) or remotely (e.g., in a \textit{distributed database}).

\begin{figure}[!]
\centering
\scriptsize
\includegraphics[width=0.6\textwidth]{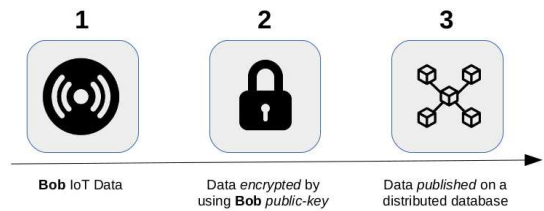}
\caption{$Data\ Encryption\ Process$}
\label{EncryptionDataGraph}
\end{figure}

\textit{}\\
\textbf{Data Encryption}: The data encryption is performed by using the \textit{tracker} device \textit{public key}.
In this way only it has the possibility to decrypt the data by using its \textit{private key}, although the involved \textit{entity} has the access to that data in encrypted form. 
The entire process has been summarized in Figure~\ref{EncryptionDataGraph}.

How already happens in the context of the \textit{blockchain} technology, where the private key cryptography mechanism provides a powerful ownership method that fulfills the authentication requirements (i.e., the ownership is \textit{private-key-based}), without the need to share more personal information, also in this context such a mechanism grants both \textit{privacy} and \textit{ownership}.

When there is the need to investigate about an \textit{entity} by using such encrypted data, for instance in case of a criminal event, such as a kidnapping or a theft, the data access can be obtained through the involved authorities in charge.
In case of minor events, it is possible to exclude this information, using the other ones (e.g., \textit{location}, \textit{time-stamp}, etc.).

\textit{}\\
\textbf{Data Hashing}: The connection between the encrypted data, stored locally or remotely, and the entity is possible by using as data-name a string generated by a \textit{hash function}~\cite{bakhtiari1995cryptographic}. 
Such a function is a special class of \textit{hash} functions largely used in cryptography.
Some common examples are: \textit{MD4}~\cite{rivest1992md4}, \textit{SHA}~\cite{bellare1994optimal}, \textit{TIGER}~\cite{mendel2007cryptanalysis}, and \textit{WHIRLPOOL}~\cite{stallings2006whirlpool}.

\begin{figure}[!]
\centering
\scriptsize
\includegraphics[width=0.6\textwidth]{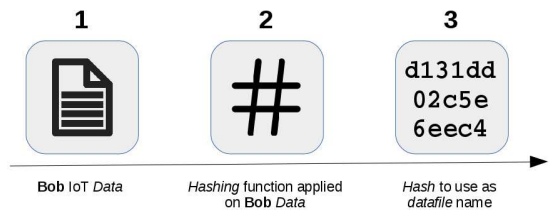}
\caption{$Data\ Hashing\ Process$}
\label{HashWorkingModelGraph}
\end{figure}

In more detail, by adopting a mathematical algorithm is possible to map data (characterized by arbitrary size) to a bit string (characterized by a fixed size).
The result is defined \textit{hash} and it represents a one-way function that is infeasible to invert. 
The literature usually refers to the input data as \textit{message} and to the output data (i.e., the \textit{hash}) as \textit{message digest} or \textit{digest}.
 
Through an \textit{hash} process, whose process is shown in Figure~\ref{HashWorkingModelGraph}, is possible to validate the data integrity of a file, detecting all modification since each of them changes the \textit{hash} output.
While an encryption process represents a \textit{two-way function} based on the \textit{encryption} and \textit{decryption} operations, hashing represents a \textit{one-way function} that transforms in an irreversible manner the source \textit{data} used as input into a plain text output (i.e., the \textit{hash} of \textit{data}).

\subsection{IoE Technology Spread}
\label{IoeTechnologySpreadSubsection}
As happened with other similar technologies, even in the case of the proposed \textit{IoE} one, the greatest obstacle to overcome is the spread across users of such a technology.

Although it is possible to create a new network of devices that operate according to the proposed \textit{IoE}  paradigm, we can substantially reduce this problem by integrating the \textit{IoE} network into the existing \textit{wireless-based} ones (e.g., \textit{IoT} and \textit{mobile}).
This process, which allows us to maximize the \emph{IoE} potential, can be facilitate by adopting several strategies, such as, the following ones:

\begin{enumerate}[(i)]
\item designing simple and transparent procedure of integration of the needed \textit{IoE} functionalities in the existing \textit{tracker} devices, for instance, by integrating these as a \textit{service} in the new devices, by recurring to a simple and well documented firmware/software upgrade process, or by making available an application, in those cases where the \textit{trackers} or the \textit{entities} are implemented in devices that allow us this solution (e.g., \textit{smart-phones}, \textit{tablets}, etc.);
\item making effective campaigns of information aimed to underline the advantages for each user that joins the \textit{IoE} network, empathizing the gained opportunity to exchange information between a large community of users, an huge amount of valuable data that they can exploit in many contexts, such as that of \textit{security} taken into account in this paper;
\item offering benefits to the users that join their devices to the \textit{IoE} network as \textit{trackers}, allowing the system to perform the \textit{entity detection} and the \textit{distributed-ledger registration} tasks.
Such a benefits could include the free-use of some services related to the \textit{IoE} network, such as, for instance, the services used for the remote data storage.
\end{enumerate} 

As previously underlined, the exploitation of the mobile network contributes to impress a substantial acceleration to the spread of the \textit{IoE} network, since such a network already involves an enormous number of devices that are potentially configurable, by recurring to simple applications, to operate according to the \textit{IoE} paradigm.
In this case, the information related to the geographic \textit{location} of the \textit{trackers} can be obtained by a local service (i.e., \textit{GPS}) or by querying the mobile \textit{cell} to which the \textit{tracker} is connected.
The sensors data related to the \textit{tracker}-side will be those available for that device, otherwise this kind of data will be absent.

A consideration should be made about the fact that the use case taken into account in this paper is based on the interaction between \textit{entities} and \textit{trackers}, implementing by using custom (e.g., wearable solutions) or standard (\textit{IoT}, \textit{smart-phone}, and \textit{tablet}) devices, but the \textit{IoE} potentiality could be improved by adding to the \textit{IoE} network other classes of devices such as, for instance, \textit{routers}, \textit{access-points}, \textit{hot-spots}, and many others.
Although this type of expansion is potentially practicable, it requires an implementation effort that is greater than that required by using the devices we considered in this paper.

\textit{}\\
\textbf{Business Models}: Some conclusive general observations are about the exploitation of the proposed \textit{IoE} paradigm in the context of a hypothetical commercial scenario.
From the point of view of a \textit{Business-to-Business} (\textit{B2B}) model, we can start by observing that many financial analysts underline that only the area related to the \textit{IoT} has given rise to an interesting and profitable financial market, whose value in the next \textit{5-10} years has been estimated around trillions of dollars~\cite{kranz2017industrial}.

Consequently, as specialized sub-area of the \textit{wireless-based} technologies market, the proposed \textit{IoE} paradigm could offer new stimulating and profitable opportunities, considering that its applications involve a huge number of customers, both private and commercial ones.
Summarizing, the activity core could be oriented towards the development of \textit{IoE} solutions for business customers, who in turn can offer this service to their customers, according to a \textit{Business-to-Consumer} (\textit{B2C}) model.

Such solutions involve both hardware and software aspects, from the hardware/software development of the \textit{IoE} devices (e.g., \textit{wearable devices}, \textit{smart-phone applications}, \textit{vehicle equipments}, etc.) to the management of the needed services (e.g., \textit{unique identifier distribution}, \textit{remote storage}, etc).

In some cases, these opportunities could be further expanded by defining and offering services in partnership with public and/or private investigative agencies (e.g., \textit{security guards}, \textit{local police}, etc), giving rise to a very interesting transversal market.

A \textit{B2C} scenario could also include other services such as, for instance, the management of \textit{entities} initially directly managed by customers or the development and commercialization of custom hardware and software solutions.

\section{Conclusion}
\label{ConclusionSection}
In this \textit{Internet}-based age, the enormous benefits related to the new technologies are dramatically jeopardized by a series of security issues given by an ever increasing number of people that try to get advantages from them, in a fraudulent way.
This scenario of insecurity is further complicated by the traditional security issues that affect our modern societies, such as, for instance, kidnappings, frauds, thefts, and so on.

The state-of-the-art security paradigms do not exploit in a better way the opportunities offered by some powerful technologies such as those related to the \textit{wireless-based} smart devices or the \textit{Internet of Things}, which involves millions of active devices, or those related to the \textit{blockchain-based distributed ledgers}, which allow to certify a series of events.

This paper introduces a new security paradigm, which we baptized \textit{Internet of Entities} (\textit{IoE}), designed to join the capabilities offered by the \textit{wireless-based} devices environment with the certification capability offered by the \textit{blockchain-based distributed ledgers}.
It is mainly based on two core components, \textit{entities} and \textit{trackers}, which are billion of new or already-existing devices able to operate interchangeably across the \textit{IoE} environment.

Although the proposed paradigm is based on existing and wide spread technologies, it offers a novel way to trace in a certified and anonymous way the activity of an \textit{entity}, \textit{person} or \textit{object}, exploiting a combination of \textit{wireless-based} and \textit{blockchain-based} technologies, which produce valuable, exploitable, and investigative-valid data.

The same mechanisms adopted in the \textit{blockchain-based} applications have been exploited in the proposed paradigm in order to ensure the \textit{immutability} of data remotely stored on a \textit{blockchain-based} distribute ledger, as well as their \textit{anonymity}.

The concept of \textit{robust network in its unstructured simplicity}, expressed by \textit{Satoshi Nakamoto} during his \textit{Bitcoin} formulation~\cite{nakamoto2008bitcoin}, well describes also the \textit{Internet of Entities} network, whose capabilities are destined to grow, day after day, thanks to the continuous introduction of new \textit{wireless-based} devices, which provide an ever expanding \textit{IoE} coverage area.

Concluding, if on the one hand, the proposed $IoE$ paradigm can be easily implemented by exploiting existing and wide spread technologies and infrastructures, on the other hand, it produces a series of advantages for the community, revealing a great potential for growth in many real-world scenarios, such as that of the security taken into consideration in this paper.

\bibliographystyle{splncs03}
\bibliography{paper} 

\end{document}